\documentclass{aa}
\usepackage{graphicx}

\def\ltsim{\raise 2pt \hbox {$<$} \kern-1.1em \lower 4pt \hbox {$\sim$}}
\def\ltapprox{\raise 2pt \hbox {$<$} \kern-1.1em \lower 5pt \hbox {$\approx
$}}
\def\gtsim{\raise 2pt \hbox {$>$} \kern-1.1em \lower 4pt \hbox {$\sim$}}
\def\gtapprox{\raise 2pt \hbox {$>$} \kern-1.1em \lower 5pt \hbox {$\approx
$}}

\begin{document}

\title{Shock acceleration as origin of the radio relic in A\,521?}
\author{S.~Giacintucci\inst{1,}\inst{2},
T.~Venturi\inst{1},
G.~Macario\inst{1,}\inst{3},
D.~Dallacasa\inst{1,}\inst{3},
G.~Brunetti\inst{1},
M.~Markevitch\inst{2},
R.~Cassano\inst{1,}\inst{3},
S.~Bardelli\inst{4},
R.~Athreya\inst{5}}

\institute
{
INAF -- Istituto di Radioastronomia, Via Gobetti 101, I--40129, Bologna, Italy 
\and
Harvard--Smithsonian Centre for Astrophysics, 
60 Garden Street, Cambridge, MA 02138, USA
\and
Dipartimento di Astronomia, Universit\`a di Bologna,
via Ranzani 1, I--40126, Bologna, Italy
\and
INAF -- Osservatorio Astronomico di Bologna, Via Ranzani 1, 
I--40126 Bologna, Italy
\and
Tata Institute of Fundamental Research,
National Centre for Radio Astrophysics,
Ganeshkhind, Pune 411 007, India}

\date{}
\date{Received 28 - 01 - 2008; accepted 22 - 03 - 2008}
\titlerunning{Shock acceleration as origin of the radio relic in A\,521?}
\authorrunning{Giacintucci et al.}
\abstract
{}
{We present new high sensitivity observations of the radio relic in A\,521
carried out with the Giant Metrewave Radio Telescope at 327 MHz and with the
Very Large Array at 4.9 and 8.5 GHz.}
{We imaged the relic at these frequencies and carried out a detailed spectral
analysis, based on the integrated radio spectrum between 235 MHz and 4.9 GHz, 
and on the spectral index image in the frequency range 327--610 MHz. To this 
aim we used the new GMRT observations and other proprietary as well as archival 
data. We also searched for a possible shock front co--located with the relic on a 
short archival Chandra X--ray observation of the cluster.}
{The integrated spectrum of the relic is consistent with a single power law; 
the spectral index image shows a clear trend of steepening going from the 
outer portion of the relic toward the cluster centre. We discuss the origin 
of the source in the light of the theoretical models for the formation of 
cluster radio relics. Our results on the spectral properties of the relic 
are consistent with acceleration of relativistic electrons by a shock in 
the intracluster medium. This scenario is further supported by our finding
of an X--ray surface brightness edge coincident with the outer border of
the radio relic. This edge is likely a shock front.}
{}

\keywords{radio continuum: galaxies - galaxies: clusters: general - galaxies:
clusters: individual: A\,521}

\maketitle
\section{Cluster radio relics and the case of A\,521}\label{sec:intro}

The radio emission from clusters of galaxies comes into two 
flavours. In addition to the individual radio sources associated 
with cluster galaxies, a fraction of massive and X--ray luminous 
clusters with clear signature of ongoing mergers, hosts large 
scale diffuse radio sources, known as {\it radio halos}, if located 
at the cluster centre, and {\it relics}, if located in peripheral 
regions. The synchrotron emission in these diffuse sources arises 
directly within the intracluster medium (ICM), and probes the existence 
of non--thermal components spread over the cluster scale (e.g.,
the review paper by Feretti 2005).

Particle acceleration via turbulence injected in the cluster volume 
during mergers represents a promising possibility to understand the 
origin of radio halos (see the review papers by Brunetti 2003, 2004; 
Sarazin 2004; Petrosian \& Bykov 2008). Recent statistical analysis 
(e.g., Kuo, Hwang \& Ip 2004; Cassano \& Brunetti \cite{cb05}; Cassano, Brunetti 
\& Setti 2006; Cassano et al. 2007) and deep radio observations of 
samples of galaxy clusters (Brunetti et al. \cite{brunetti07}; Venturi 
et al. \cite{venturi07}; Venturi et al. 2008) have provided 
further support to this scenario. On the other hand, our understanding 
of the origin of relics is still limited. Only a handful of radio relics 
have been observed in some detail (e.g., A\,2256, Clarke \& En\ss{}lin 
\cite{clarke06}; A\,3667, R\"ottgering et al. 1997; A\,2744, Orr\'u et 
al. \cite{orru07}). Integrated radio spectra over a wide range of 
frequencies are available for few relics only (e.g., 1257+275 in the Coma 
cluster; Andernach et al. 1984, Thierbach, Klein \& Wielebinski 
\cite{thierbac03} and references therein). All models proposed so far for 
the relic formation invoke the presence of a shock within the X--ray gas 
(En\ss{}lin et al. \cite{ensslin98}; Roettiger et al. 1999; En\ss{}lin 
\& Gopal--Krishna \cite{ensslin01}; Hoeft \& Br\"uggen 2007).
\\
\\
Here we focus our attention on the radio relic in A\,521 (Ferrari et al. 
2006; Giacintucci et al. 2006, hereinafter GVB06), an X--ray luminous and 
massive galaxy cluster (L$_{\rm X \, [0.1-2.4\, keV]} \sim 8 \times 10^{44}$ 
erg s$^{-1}$; virial mass M$_{\rm v} \sim 1.9 \times 10^{15}$ 
M$_{\odot}$\footnote{from the correlation between the X--ray luminosity 
and M$_{\rm v}$ reported in Cassano, Brunetti \& Setti (\cite{cbs06})}) at  
redshift z=0.247. Multiple merging episodes are known to be occurring in 
this very disturbed cluster, whose properties are indicative of an object 
in a complex dynamical state, and still accreting a number of smaller 
mass concentrations (e.g., Maurogordato et al. \cite{maurogordato00}; Ferrari 
et al. \cite{ferrari03b} and 2006; see also Fig.~1 in GVB06 for a sketch of 
the multiple optical and X--ray substructures in the cluster).

%
%

\begin{table*}[t]
\caption[]{Summary of the radio observations.}
\begin{center}
\footnotesize
\begin{tabular}{lcccclccc}
\hline\noalign{\smallskip} 
Telescope & RA$_{J2000}$& DEC$_{J2000}$  & Observation  & $\nu$ & $\Delta \nu$  & t  & HPBW, p.a.  &   rms      \\ 
          & (h, m, s)  & ($^\circ$, $^{\prime}$,$^{\prime \prime}$)  &  date    &   MHz      &  MHz          &  min       & (full array , $^{\prime \prime} \times^{\prime \prime}$, $^{\circ}$)&   $\mu$Jy b$^{-1}$\\
\noalign{\smallskip}
\hline\noalign{\smallskip}
GMRT       & 04 54 09.00 & $-$10 14 19.0 & Nov 06 & 327    &  16 $\diamond$ &  330  & 10.6$\times$9.6, $-$19   & 90--100  $*$\\
VLA--BnA   & 04 54 16.28 & $-$10 16 05.9 & Jun 06 & 4860   &  50  & 110   & 1.0$\times$0.8, $-$68 & 10  \\	
VLA--CnB   & 04 54 16.28 &$-$10 16 05.9& Oct 06 & 4860   &  50  &  60   & 4.0$\times$2.0, $-$82 & 15   \\
VLA--BnA   & 04 54 16.28 & $-$10 16 05.9& Jun 06 & 8430   &  50  &  170  & 0.6$\times$0.5, 82      &  8 \\  
 \\
\noalign{\smallskip}
\hline\noalign{\smallskip}
\end{tabular}
\end{center}
$\diamond$ the observations were performed using a total bandwidth 
of 32 MHz (USB+LSB), but only the USB dataset was used for the analysis 
(see Sec. 2.1); 

$*$ the rms noise level in the region of the relic is $\sim$ 90 $\mu$Jy b$^{-1}$, 
and increases up to 100 $\sim$ $\mu$Jy b$^{-1}$ in the outer parts of the A\,521 
field due to the presence of strong radio sources.
\label{tab:obs}
\end{table*}
%
%

A radio study of A\,521 based on 610 MHz Giant Metrewave Radio Telescope 
(GMRT) observations (GVB06) shows that the relic is a diffuse elongated 
structure located in the south--eastern periphery of A\,521, at the edge 
of a dynamically active region where galaxy group infalls into the main 
cluster are taking place (Maurogordato et al. \cite{maurogordato00}; 
Ferrari et al. \cite{ferrari03b}). The relic is at the boundary of the 
X--ray emission from the intracluster gas, at a projected distance of $\sim$ 
930 kpc from the cluster X--ray centre, and it is apparently connected 
to the most powerful radio galaxy in A\,521 (J0454--1016a) by a faint bridge 
of radio emission. Even though projection effects should be taken into account, 
this situation is similar to what is observed in the Coma cluster, where a 
bridge of radio emission connects the tails of the radio galaxy NGC\,4789 to 
the prototype relic source 1253+275 (Giovannini, Feretti \& Stanghellini \cite{giovannini91}).
\\
\\
In this paper we present a study on the origin of the relic based on new GMRT 
327 MHz high sensitivity observations of A\,521 and new high frequency and high 
resolution images of the relic region obtained from Very Large Array (VLA) 
observations at 4.9 and 8.5 GHz. The amount of radio information available 
allows us to study in detail the spectral properties of the source (integrated 
and point--to--point) over a relatively wide range of frequencies, and compare 
them to the expectations from models for its origin. The paper is organised as 
follows: Sec.~\ref{sec:obs} describes the radio observations and data reduction; 
the new GMRT 327 MHz images of the A\,521 field and relic region are presented 
in Sec.~\ref{sec:relic327}; in Sec.~\ref{sec:highfreq} we report on the analysis 
of the VLA 4.9 and 8.5 GHz images; Sec.~\ref{sec:spix} deals with the study 
of the spectral index image; the source integrated radio spectrum is 
analysed in Sec.~\ref{sec:spectrum}; the proposed scenarios for the relic 
origin are discussed in Sec.~\ref{sec:disc}; finally a summary of our results 
is given in Sec.~\ref{sec:summary}.
\\
\\
\indent We adopt the $\Lambda$CDM cosmology with H$_0$=70 km s$^{-1}$ Mpc$^{-1}$, $\Omega_m=0.3$ and $\Omega_{\Lambda}=0.7$. At the redshift of A\,521 this cosmology 
leads to a linear scale of 1$^{\prime \prime}$ = 3.87 kpc. The spectral index 
$\alpha$ is defined according to S$\propto \nu^{-\alpha}$. 

%
%
\begin{figure*}[t]
\centering
\includegraphics[angle=0,width=15cm]{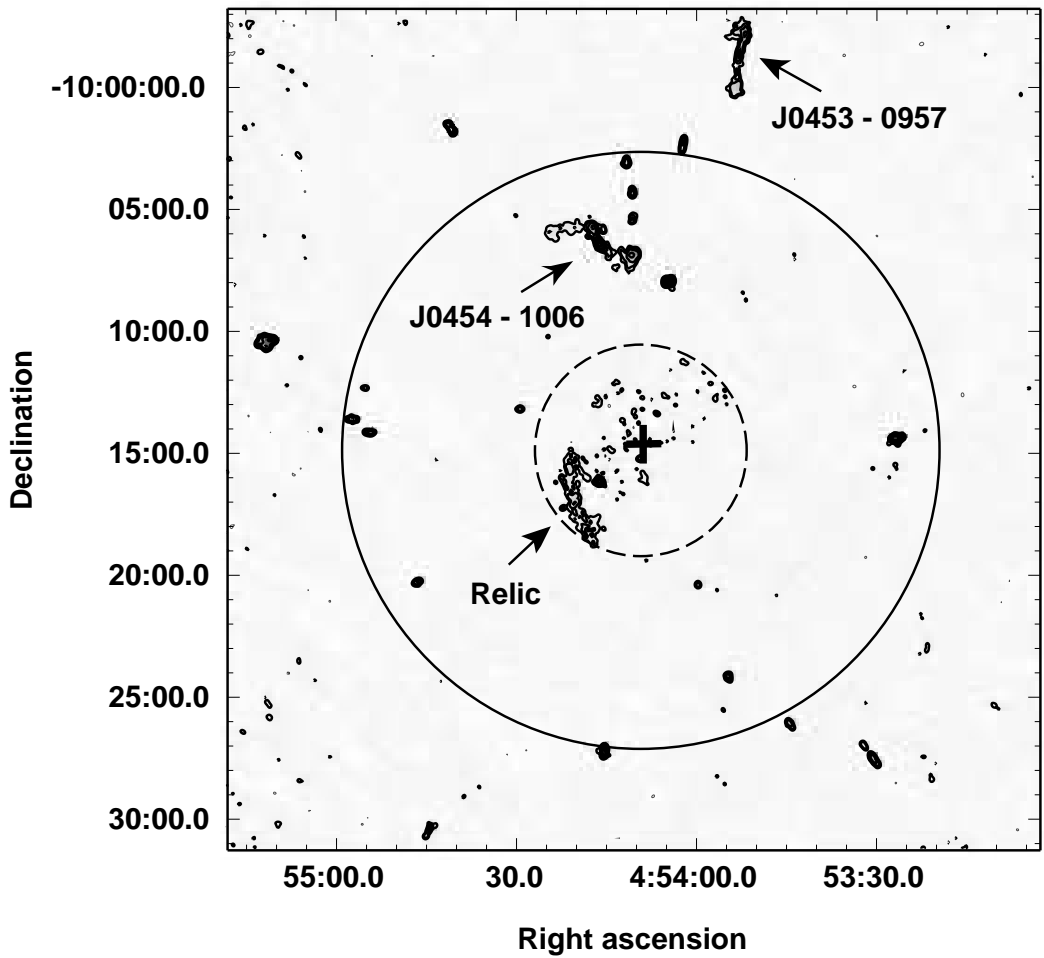}
\caption{GMRT 327 MHz contours of the $30^{\prime} \times
30^{\prime}$ region containing A\,521. Contour levels are 
spaced by a factor 2 starting from $5\sigma=\pm 0.5$ mJy b$^{-1}$. The
resolution is $10.6^{\prime \prime} \times 9.6^{\prime \prime}$,
in p.a. $-19^{\circ}$. The cross marks the X--ray centre of the 
cluster (Arnaud et al. 2000). The radius of the 
solid circle corresponds to the cluster virial radius 
$\rm R_v=2.8$ Mpc (GVB06). The dashed circle has a radius 
of $\sim 1$ Mpc, and indicates the region covered by the 
optical analysis in Ferrari et al. (\cite{ferrari03b}).}
\label{fig:327field}
\end{figure*}
%
%

\section{Radio observations and data reduction}\label{sec:obs}

We carried out high sensitivity observations of A\,521 using the GMRT at 327 MHz. 
In order to image the sky region close to the relic at higher frequency and 
resolution, and resolve the inner structure of the radio galaxy J0454--1016a, 
we performed VLA observations at 4.9 GHz in the hybrid BnA and CnB configurations, 
and at 8.5 GHz in the BnA configuration. In Table \ref{tab:obs} we summarise the 
details on all the radio observations presented in this paper. Columns in the table 
provide the following information: telescope/array; J2000 coordinates of the pointing 
centre; observing day; frequency; total bandwidth; total time on source; half 
power bandwidth (HPBW) and rms level (1$\sigma$) in the full resolution image.

\subsection{GMRT observations at 327 MHz}\label{sec:gmrt_obs}

A\,521 was observed using the GMRT at 327 MHz in November 2006 for a total integration 
time of 5.5 hours (Tab.~\ref{tab:obs}). The observations were performed using both 
the upper and lower side band (USB and LSB, respectively) for a total observing 
bandwidth of 32 MHz. The data were collected in spectral--line mode with 128 
channels/band, and a spectral resolution of 125 kHz/channel. 
\\
The USB and LSB datasets were calibrated and analysed individually using the 
NRAO Astronomical Image Processing System (AIPS) package. The bandpass calibration was 
performed using the flux density calibrator. A RFI--free channel was chosen to 
normalise the bandpass for each antenna. The calibration solutions were applied 
to the data by running the AIPS task FLGIT, which subtracts a continuum from the 
channels in the u--v plane, determined on the basis of the bandpass shape 
and using a specified set of channels. The data whose residuals exceed a chosen 
threshold are then flagged. Despite this flagging procedure, both the USB and LSB 
datasets were still affected by strong residual radio frequency interferences 
(RFI). Hence, a very accurate editing of the visibility data was carried out in 
order to identify and remove those data affected by RFI.   

In order to find a compromise between the size of the 
dataset and the need to minimize bandwidth smearing 
effects within the primary beam, the central channels 
were averaged to 6 channels of $\sim$2 MHz each after 
bandpass calibration. Given the large field of view of 
the GMRT, in each step of the data reduction we implemented 
the wide--field imaging technique to minimise the errors 
due to the non--planar nature of the sky. We used 25 facets 
covering a total field of view of $\sim 2.7 \times 2.7$ 
square degrees. After a number of phase self--calibration cycles, 
the final USB and LSB datasets were further averaged 
from 6 channels to 1 single channel\footnote{Bandwidth smearing 
is relevant only at the outskirts of the wide field, and 
does not significantly affect the region presented and 
analysed in this paper}. 

Due to residual phase errors in the LSB dataset, the USB--LSB data 
combination led to images with a quality worse than those obtained 
from the USB alone. For this reason only the USB dataset was 
used for the analysis presented in this present paper. 
A very high sensitivity (1$\sigma$) was achieved in our 
final full resolution image (Tab.~\ref{tab:obs}): 
from $\sim$ 90 $\mu$Jy b$^{-1}$ in the region of the 
relic to $\sim$ 100 $\mu$Jy b$^{-1}$ in the outer parts 
of the A\,521 field, where the quality of the image is 
limited by the presence of strong radio sources. The residual 
amplitude errors are of the order of $\ltsim$ 5\%.

%
%
\begin{figure*}[t]
\centering
\includegraphics[angle=0,width=8.5cm]{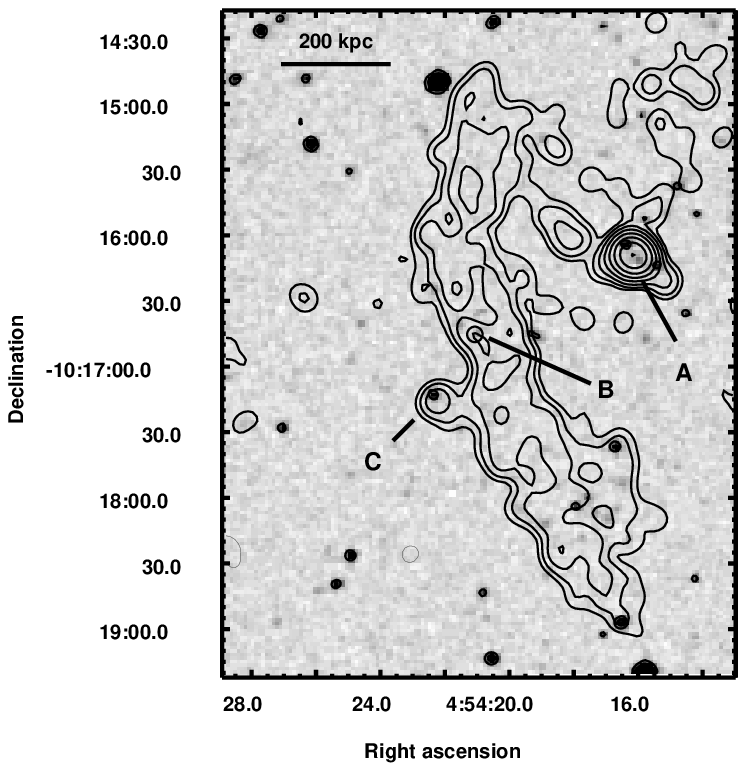}
\includegraphics[angle=0,width=8.5cm]{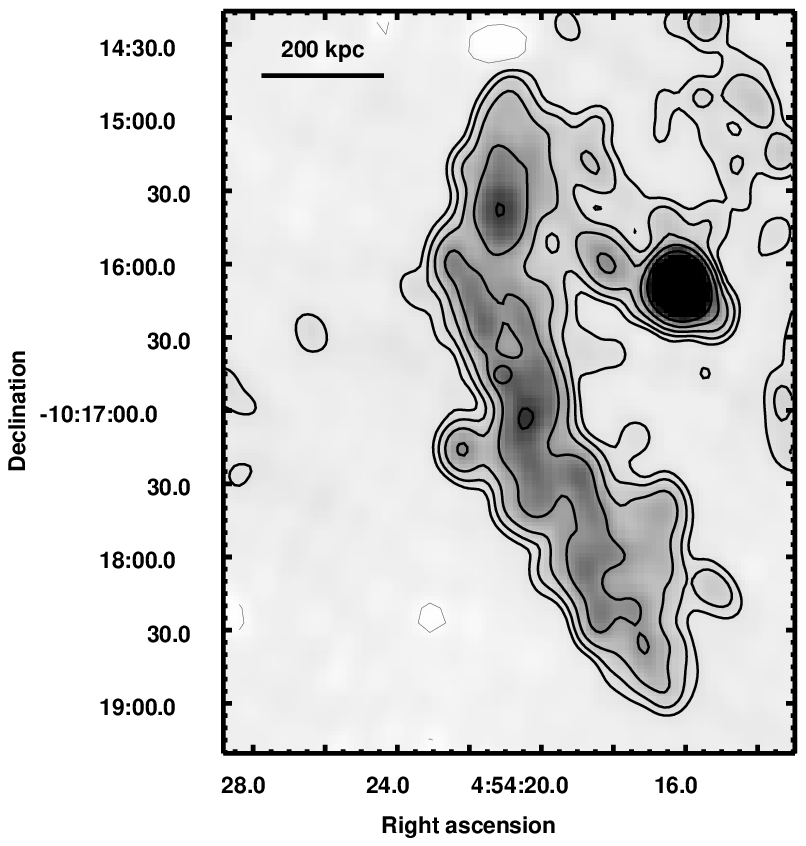}
\caption{{\bf Left panel:} full resolution GMRT contours at 327 MHz 
of the relic, overlaid on the red POSS--2 optical image. The resolution 
is $10.6^{\prime \prime} \times 9.6^{\prime \prime}$, in p.a. $-19^{\circ}$. 
Contours are spaced by a factor 2 starting from $\pm 3\sigma=0.27$ mJy 
b$^{-1}$. A, B and C indicate the position of 
the radio galaxies embedded in the relic emission. {\bf Right panel:} low 
resolution GMRT image at 327 MHz of the relic (contours and grey scale). 
The resolution is $15.0^{\prime \prime} \times 12.0^{\prime \prime}$, in p.a.
0$^{\circ}$. Contours are spaced by a factor 2 starting 
from $\pm 3\sigma=0.27$ mJy b$^{-1}$. }
\label{fig:relic327}
\end{figure*}
%
%

\subsection{VLA observations at 4.9 and 8.5 GHz}\label{sec:vla_obs}

The 4.9 GHz observations of the relic region were carried 
out using the VLA in the hybrid BnA and CnB configurations. 
The 8.5 GHz observations were performed in the BnA configuration 
(see Tab.~\ref{tab:obs} for details). A bandwidth of 50 MHz was 
used for each of the two IF channels at each frequency. All 
observations included full polarisation information. 
The data were calibrated and reduced using the pilot semi--automatic 
pipeline for the VLA data processing, a facility recently developed 
at NRAO and implemented in the AIPS package. The pipeline provided 
high quality calibrated datasets at each frequency. These latter 
were further phase self--calibrated in AIPS, in order to correct for 
residual phase variations, and used to produce the final images. 
The rms noise level achieved in the final images are reported 
in Tab.~\ref{tab:obs}. The average residual amplitude errors in the 
data are of the order of $\, \ltsim$ 5 $\%$ both at 4.9 and 8.5 GHz.

\section{A\,521 at 327 MHz}\label{sec:relic327}

\subsection{The field}

In Fig.~\ref{fig:327field} we present the GMRT full resolution 
image at 327 MHz covering the region within the cluster virial 
radius (R$\rm_{V}$=2.78 Mpc; GVB06), which is delimited by the 
solid circle. The figure shows the same $\sim 30^{\prime} \times 
30^{\prime}$ field presented at 610 MHz in Fig.~2 of GVB06. The cross 
marks the centre of the cluster X--ray emission as detected by 
ROSAT HRI (Arnaud et al. \cite{arnau00}). The dashed circle has a 
radius corresponding to $\sim$ 1 Mpc, and indicates the area covered 
by the analysis of the optical substructures in Ferrari et al. 
(\cite{ferrari03b}). The radio contours are plotted starting from 
$\pm$ 0.5 mJy b$^{-1}$, which corresponds to the 5$\sigma$ level 
in the region with the highest noise.

Three very extended radio sources are visible in the image: the radio 
relic, in the south--eastern outskirts of the cluster, and the two 
radio galaxies J0453--0957 and J0454--1006, located North of A\,521, 
and analysed at 610 MHz in GVB06. 

In addition to discrete point sources, positive residuals of 
radio emission are detected within the dashed circle 
in Fig.~\ref{fig:327field}, suggesting the presence of diffuse emission 
at the cluster centre. The investigation of this point is beyond the 
purpose of the present paper, and will be addressed in a forthcoming paper 
(Brunetti et al. to be submitted).

\subsection{The radio relic at 327 MHz}

As observed at higher frequencies, the 327 MHz radio emission within 
the inner $\sim$ 1 Mpc radius is clearly dominated by the relic.
Fig.~\ref{fig:relic327} zooms into the 327 MHz image of the
relic. In the left panel we show the full resolution contours
overlaid on the optical POSS--2 frame (grey scale). Labels A, B and 
C indicate the position of the radio galaxies embedded in the diffuse 
relic emission, and optically identified in GVB06. In the right 
panel we show an image at the resolution of $15.0^{\prime \prime} 
\times 12.0^{\prime \prime}$ with grey scale and contours overlaid 
in order to better highlight the distribution of the radio 
surface brightness across the source. The relic exhibits 
a highly elongated and arc--shaped structure with an
angular size of $\sim 4.3^{\prime}$, which corresponds to a 
linear size of $\sim 1$ Mpc. The overall morphology and 
total extent in Fig.~\ref{fig:relic327} are in good agreement 
with the images at 610 MHz (GVB06; also reported in Fig.~5) 
and at 1.4 GHz (Ferrari et al. 2006) of similar resolution. 
The relic emission along the minor axis appears on the average 
slightly wider in the 327 MHz image ($\sim 1.0^{\prime}$, i.e.,
$\sim 230$ kpc) than at higher frequencies ($\sim 200$ kpc at 610 
MHz and $\sim$160 kpc at 1.4 GHz).

\section{The relic region at 4.9 and 8.5 GHz}\label{sec:highfreq}

\subsection{The radio galaxy J0454--1016a}\label{sec:a521_j0454}

%
%
\begin{figure*}[t]
\centering
\includegraphics[width=14cm]{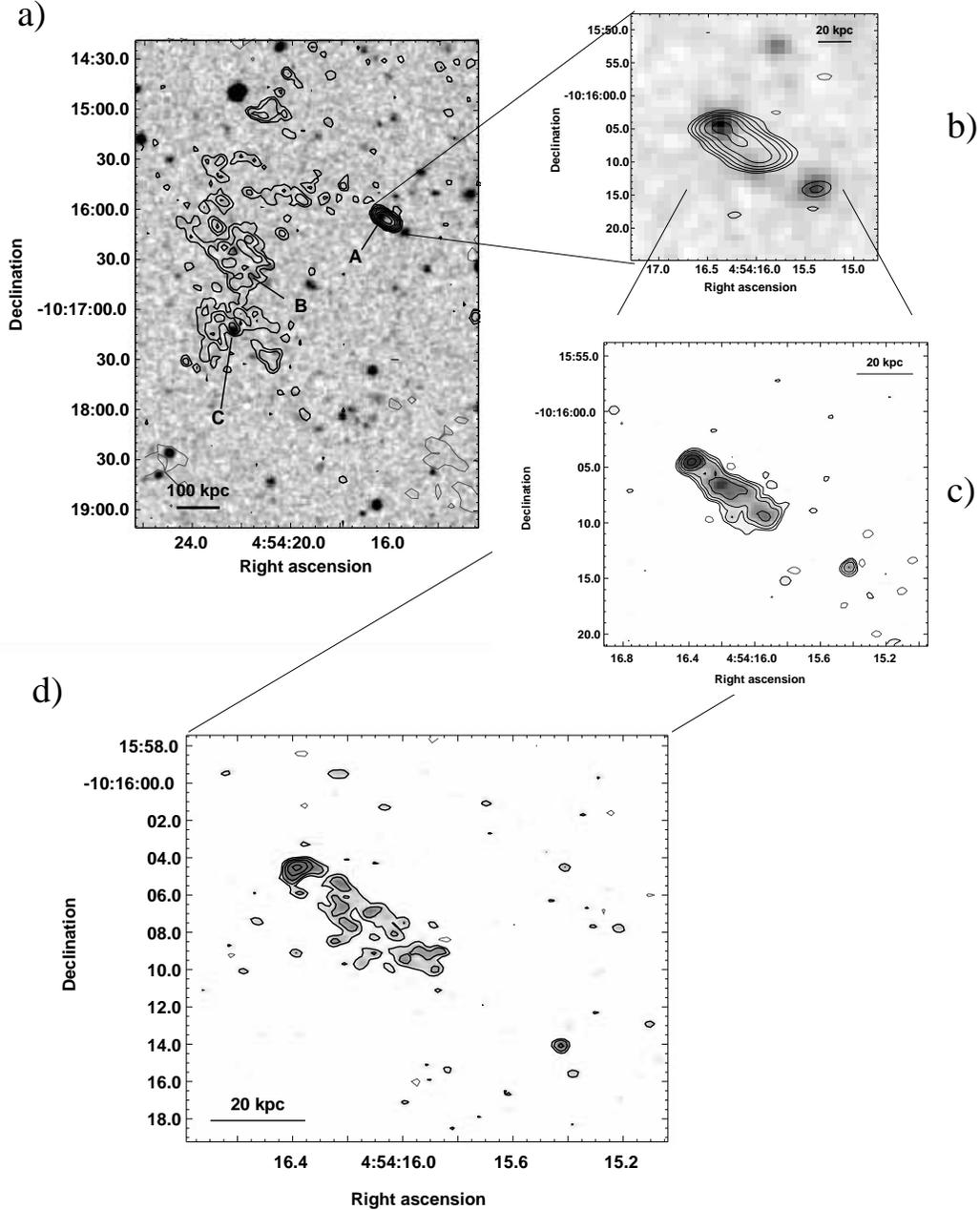}
\caption{{\bf Panel a)} -- GMRT 610 MHz full resolution contours of the
 relic on the POSS--2 image. The HPBW is $5.0^{\prime \prime} 
\times 4.0^{\prime \prime}$. The 1$\sigma$ noise level is 
40 $\mu$Jy b$^{-1}$. A, B and C indicate the position of radio 
galaxies embedded in the relic emission. {\bf Panel b)} -- VLA--CnB 4.9 GHz 
contours of J0454--1016a on the POSS--2; HPBW=$4.0^{\prime \prime} 
\times 2.0^{\prime \prime}$; 1$\sigma$=15 $\mu$Jy b$^{-1}$. {\bf Panel c)}
--  VLA--BnA 4.9 GHz image (contours and grey scale) 
of J0454--1016a; HPBW=$1.0^{\prime \prime} 
\times 0.8^{\prime \prime}$; 1$\sigma$=10 $\mu$Jy b$^{-1}$.
{\bf Panel d)} -- VLA--BnA 8.4 GHz image (contours and grey scale) 
of J0454--1016a. HPBW=$0.6^{\prime \prime} 
\times 0.5^{\prime \prime}$; 1$\sigma$=8 $\mu$Jy b$^{-1}$.
In all the panels the contour levels are spaced by a factor 
2 starting from $\pm$3$\sigma$.} 
\label{fig:a521_compo}
\end{figure*}
%
%

Source A in Fig.~\ref{fig:relic327} is the cluster radio galaxy 
J0454--1016a. A faint bridge of emission is clearly detected 
between the northern part of the relic and this radio galaxy, which is 
located at $\sim 350$ kpc (projected) from the relic. This emission 
was also observed at 610 MHz, and suggested the presence of a physical 
link between the two objects (GVB06). We investigated the possible connection
between the relic and J0454--1016a by means of VLA 4.9 and 8.5 GHz 
observations (Tab.~\ref{tab:obs}), aimed to resolve the inner 
structure of the source, and search for evidence of a connection 
with the nearby relic, for example in the form of 
bent jets and/or extended emission in that direction.

%
%
\begin{table}[t] 
\caption[]{Properties of the radio galaxy J0454--1016a.}
\begin{center}
\begin{tabular}{lc}
\hline\noalign{\smallskip}
S$_{\rm 327 \, MHz}$ (mJy) & 46.0 $\pm$ 2.3 $^{i)}$  \\
S$_{\rm 610 \, MHz}$ (mJy) & 27.7 $\pm$ 1.4 $^{ii)}$  \\
S$_{\rm 1.4 \, GHz}$ ~(mJy) & 15.0 $\pm$ 0.8 $^{iii)}$  \\
S$_{\rm 4.9 \, GHz}$ ~(mJy) & ~6.2 $\pm$ 0.3 $^{iv)}$  \\
S$_{\rm 8.5 \, GHz}$ ~(mJy) & ~3.6 $\pm$ 0.2 $^{v)}$\\
\smallskip
logP$_{\rm 1.4 \, GHz}$ (W Hz$^{-1}$)  &  24.41  \\
$\alpha_{\rm 4.9 \, GHz}^{\rm 8.5 \, GHz}$ & 0.79$\pm$0.13 \\
& \\
\hline
\end{tabular}
\end{center}
Notes to Tab.~\ref{tab:j0454}: $i)$ from Fig.~\ref{fig:relic327} (left panel); 
$ii)$ from GVB06; $iii)$ from VLA archival data (Obs. Id. AF\,0390);
$iv)$ from panel {\it b)} of Fig.~3; $v)$
from panel {\it d)} of Fig.~3.
\label{tab:j0454}
\end{table}
%

J0454--1016a is the most powerful radio galaxy in A\,521 (GVB06; 
Tab.~\ref{tab:j0454}), and is identified with the galaxy $\#143$ 
(v=74282 km s$^{-1}$, I=17.00) in the optical catalogue by
Ferrari et al. (\cite{ferrari03b}). Fig.~\ref{fig:a521_compo} 
shows the region of the relic and J0454--1016a (labelled as A) 
in increasing frequency (from 610 MHz to 8.5 GHz) and resolution 
order ($\sim 5^{\prime\prime}$ to $\sim 0.5^{\prime\prime}$)
going from panels {\it a)} to {\it d)}. Panel {\it a)} shows 
the full resolution image at 610 MHz.
The VLA--CnB 4.9 GHz image of J0454--1016a is presented in 
panel {\it b)}; the BnA full resolution images at 4.9 
and 8.5 GHz are shown in panels {\it c)} and {\it d)}, respectively. 
J0454--1016a appears extended in all the images 
with a largest linear size of $\sim 30$ kpc. Its radio structure 
is consistent with a head--tail morphology. The compact component, 
detected at 4.9 and 8.4 GHz (panels {\it c} and {\it d}), is coincident 
with the nucleus of the host galaxy.

The extended emission is entirely located South--West of the compact 
component, and there is no evidence of any emission in the direction 
of the radio bridge detected at 327 MHz (Fig.~\ref{fig:relic327}) 
and 610 MHz (panel {\it a}). This result might rule out a 
physical connection between the diffuse emission from the 
relic and this radio galaxy, whose tail extends (at least in 
projection) in a nearly opposite direction.

%
%
\begin{figure}[h]
\centering
\includegraphics[angle=0,width=8cm]{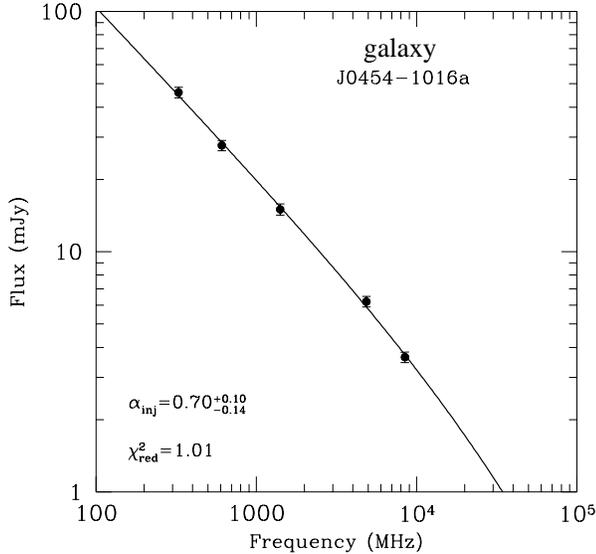}
\caption{Integrated radio spectrum of J0454--1016a between
327 MHz and 8.5 GHz. The solid line is the best fit of the
CI model. The value of $\alpha_{\rm inj}$ provided by the fit, 
along with the reduced $\chi^2$, is also reported.} 
\label{fig:j0454_spectrum}
\end{figure}
%
%

The flux densities available for J0454--1016a are collected 
Tab.~\ref{tab:j0454}, where we also report the 1.4 GHz radio 
power (from VLA archival data, Obs. Id. AF\,0390, re--analised in 
Giacintucci 2007), and the spectral index in the 327 MHz--8.5 
GHz interval. The flux density measurements on both images at 
4.9 GHz (panels {\it b} and {\it c} in Fig.~\ref{fig:a521_compo}) are 
consistent within the errors. Fig.~\ref{fig:j0454_spectrum} 
shows the integrated radio spectrum of J0454--1016a between 
327 MHz and 8.4 GHz, derived using the values in 
Tab.~\ref{tab:j0454}. 

We fitted the spectrum with the Synage++ package 
(Murgia 2001) assuming a continuous injection model 
(CI; Kardashev 1962). The best fit is shown 
as solid line in Fig.~\ref{fig:j0454_spectrum}, and 
provides an injection spectral index $\alpha_{\rm inj}= 
0.70^{+0.10}_{-0.14}$. Even though there is an indication 
of a spectral steepening above 4.9 GHz, the spectrum is consistent with 
a single power law with slope $\alpha =\alpha_{\rm inj}$.

The spectral shape in Fig.~\ref{fig:j0454_spectrum} is similar 
to what is observed in other active and low luminosity radio 
galaxies (e.g., Parma et al. \cite{parma02}).

\subsection{The radio relic at 4.9 GHz}\label{sec:relic5ghz}

%
%
\begin{figure*}[t]
\centering
\hspace{-1cm}\includegraphics[angle=0,width=15cm]{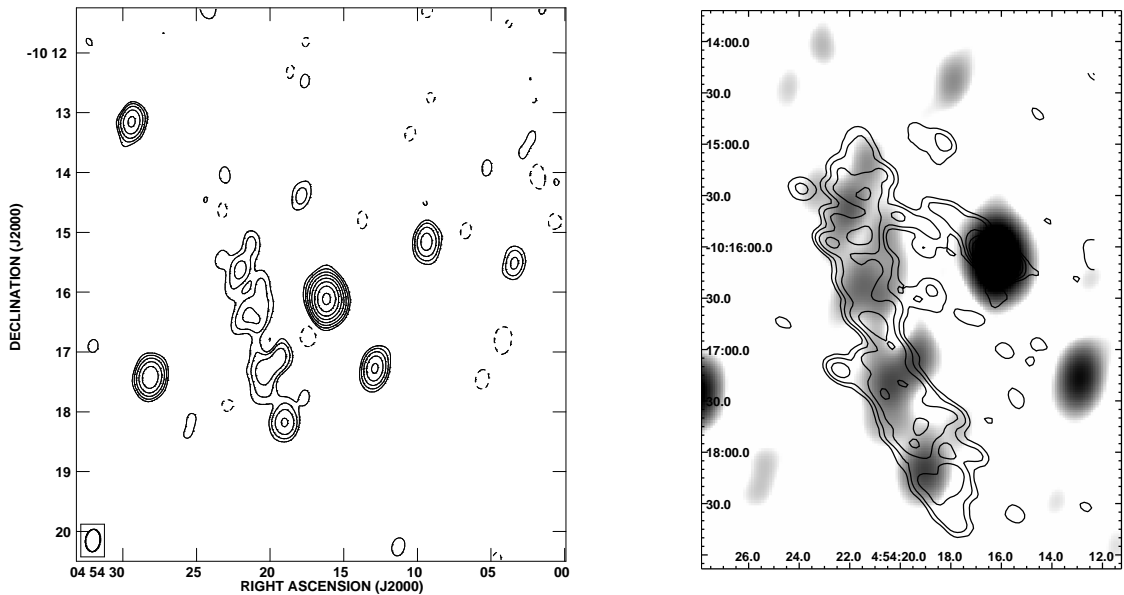}
\caption{VLA 4.9 GHz image of the radio relic as contours 
(left panel), and grey scale on the GMRT 610 MHz 
image (right panel). The resolution of the 4.9 GHz image 
is $22.0^{\prime \prime} \times 15.0^{\prime \prime}$, 
and the 1$\sigma$ noise level is 15 $\mu$Jy b$^{-1}$. The 
resolution of the 610 MHz image is $13.1^{\prime \prime} 
\times 8.1^{\prime \prime}$, and the 1$\sigma$ noise level 
is 40 $\mu$Jy b$^{-1}$. The radio contours are spaced by a 
factor 2, starting from $\pm$0.045 mJy b$^{-1}$ and 
$\pm$0.012 mJy b$^{-1}$ at 4.9 GHz (left panel) and 
610 MHz (right panel), respectively.}
\label{fig:a521_relic4.9}
\end{figure*}
%
%

The most interesting result of the VLA 4.9 GHz 
observations is the detection of the radio relic, 
shown in Fig.~\ref{fig:a521_relic4.9}. This is the 
second detection of a radio relic at a frequency as high as 
4.9 GHz, after the relic source 1253+275 in the Coma 
cluster (Andernach et al. 1984; Thierbach, Klein \& 
Wielebinski 2003). The image was obtained from the 
VLA--CnB data (Tab.~\ref{tab:obs}), 
tapered to a resolution of $22.0^{\prime \prime} 
\times 15.0^{\prime \prime}$. In the left panel the 
relic is shown as contours, while in the right panel 
the 4.9 GHz emission is reported in grey scale with 
the GMRT 610 MHz contours overlaid (GVB06). The faintest
features of the relic emission are significant at the 
3$\sigma$ level (1$\sigma$= 15 $\mu$Jy b$^{-1}$), and 
the peaks at the level of 12$\sigma$. We point out 
that, given the short observing time (Tab.~1) and 
lack of short baselines (minimum baseline $\sim$ 1 
k$\lambda$), the u--v coverage of the 4.9 GHz 
observations is not adequate to properly image 
the whole diffuse emission associated with the relic. 
For this reason the details of the structure 
of the source in Fig.~\ref{fig:a521_relic4.9} 
might be not fully reliable, and deeper 4.9 GHz 
observations with a more appropriate array and u--v
coverage are needed to better determine the relic 
morphology at this frequency.

\section{Spectral index image of the relic}\label{sec:spix}

Spectral index imaging of cluster radio relics is a 
powerful tool to investigate and understand their origin,
evolution, and connection with the merging activity of the 
hosting cluster. In particular, the spectral index images 
provide important information on the energy spectrum of the 
radio emitting electrons and magnetic field distribution in 
these sources (e.g., Clarke \& En\ss{}lin \cite{clarke06}). 

We obtained the image of the spectral index distribution 
in the A\,521 relic in the frequency range 327--610 MHz, 
by comparing the GMRT image at 327 MHz shown here (Fig.~2; 
right panel) with the GMRT 610 MHz image obtained from the 
observations presented in GVB06. The images were produced 
with the same cell size, u--v range, and restoring beam. 
For the details we refer to Tab.~\ref{tab:spix}, where we provide 
the u--v range, beam and noise level (1$\sigma$) 
of the two images. The images were aligned, the 
pixels with brightness below the 3$\sigma$ level 
were blanked, and finally the image combination 
was carried out to create the spectral index image 
using the Synage++ package. The resulting spectral index image 
is shown in Fig. \ref{fig:sp_images} (colour), with the 610 MHz 
contours overlaid.

%
\begin{table}
\caption[]{Details of the 327--610 MHz spectral index image.}
\begin{center}
\footnotesize
\begin{tabular}{lc}
\hline\noalign{\smallskip} 
Frequency range ($\nu_{1} - \nu_{2}$) &  327 MHz--610 MHz \\
u--v range & 0.12--25 k$\lambda$\\
HPBW, p.a. & 15.0$^{\prime \prime} \times$ 12.0$^{\prime \prime}, 0^{\circ}$\\ 
rms ($\nu_{1}$) & 90 $\mu$Jy b$^{-1}$ \\
rms ($\nu_{2}$) &  40 $\mu$Jy b$^{-1}$ \\
\noalign{\smallskip}
\hline\noalign{\smallskip}
\end{tabular}
\end{center}
\label{tab:spix}
\end{table}
%
%

\subsection{General features of the spectral index image}\label{sec:spix_images}

The distribution of the spectral index in Fig.~6 shows
different features along the two axis of the relic. Along 
the major axis a number of irregularities provide a 
rather patchy appearence of the spectral index image.
Such patchiness might arise from irregularities in 
the u--v coverage occurring at different spacings 
at the two frequencies. Along the minor axis a 
steepening of the spectral index from the eastern 
edge of the relic toward the western border is visible.
This trend is real and not driven by a misalignement 
between the images at 327 and 610 MHz. We carefully checked 
the correct alignment of the images using the point sources in 
the relic region: the spectral index distribution 
appears uniform in each of them and consistent with 
the value of $\alpha$ obtained from their total flux 
density at the two frequencies. This is also clear 
from Fig.~\ref{fig:sp_images}, where for example the 
radio galaxy J0454--1016a (labelled as A) has an average 
spectral index $\alpha = 0.75 \pm 0.05$, 
which is in good agreement with the source integrated 
spectrum shown in Fig.~\ref{fig:j0454_spectrum} 
(see also Tab.~2).

\subsection{Analysis of the radial steepening}\label{sec:spix_steep}

%
%
\begin{figure}[h]
\centering
\includegraphics[angle=0,width=7.5cm]{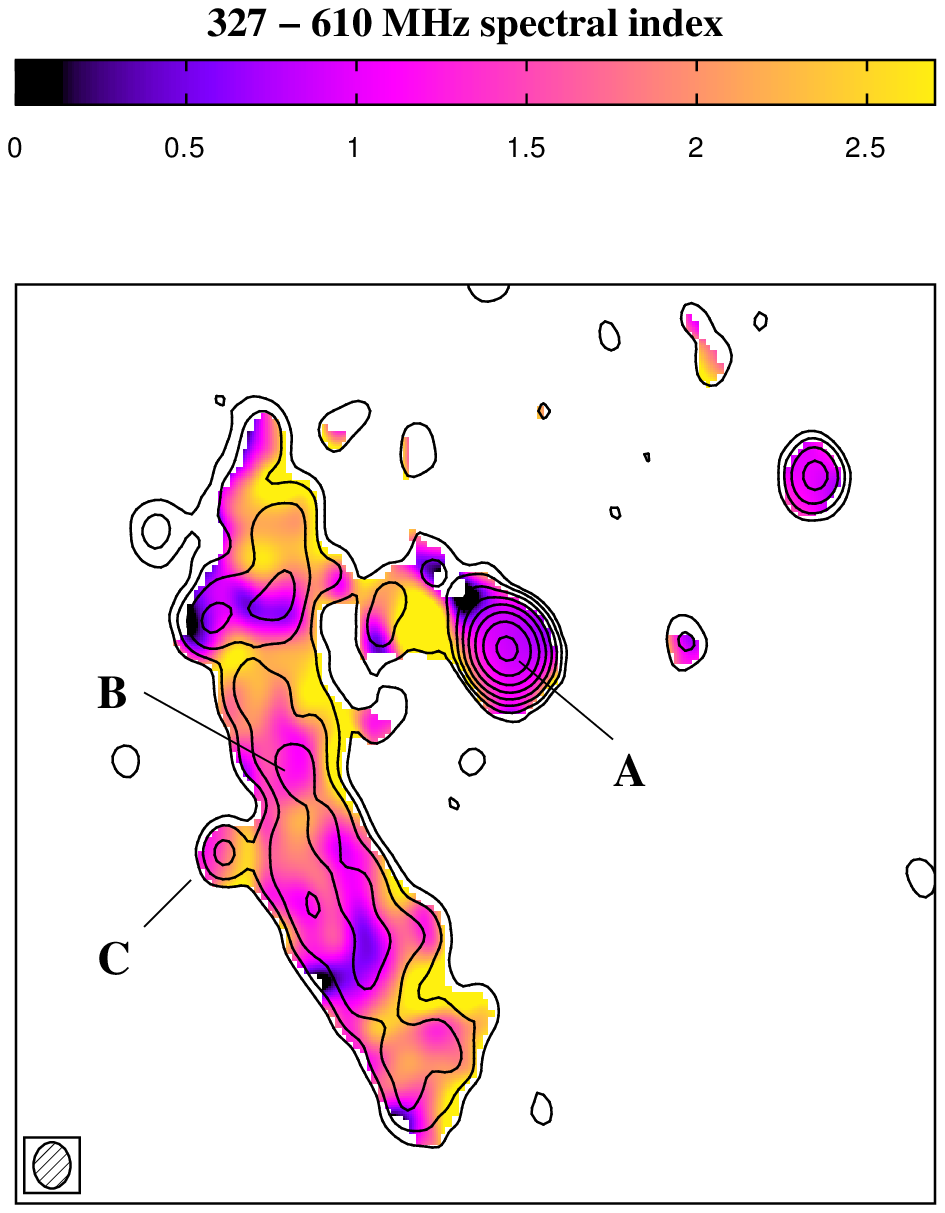}
\caption{Spectral image of the relic in A\,521 in the 327--610 MHz
range, with overlaid the GMRT 610 MHz contours at levels 
0.12, 0.24, 0.48, 0.96, 2, 4 mJy b$^{-1}$. In both images (spectral index 
and total intensity) the resolution is $15^{\prime\prime} 
\times 12^{\prime\prime}$, p.a. 0$^{\circ}$, as shown by the 
ellipse in the upper--left corner of the image. A,B and C indicate 
the position of radio galaxies embedded in the relic emission
(see Figs.~2 and 3).}
\label{fig:sp_images}
\end{figure}
%
%

%
%
\begin{figure*}[t]
\centering
\includegraphics[angle=0,width=7cm]{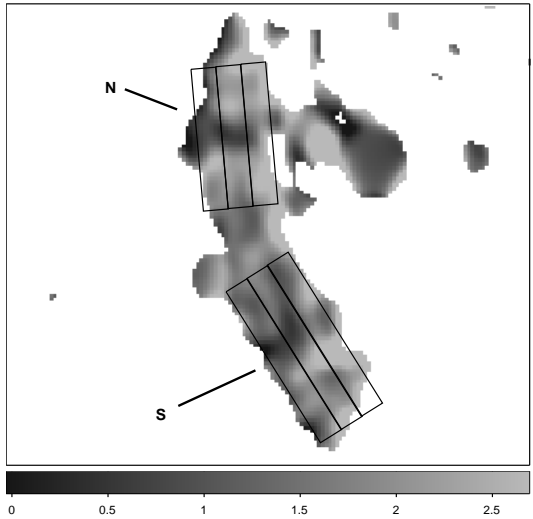}
\hspace{1.5cm}
\includegraphics[angle=0,width=7.3cm]{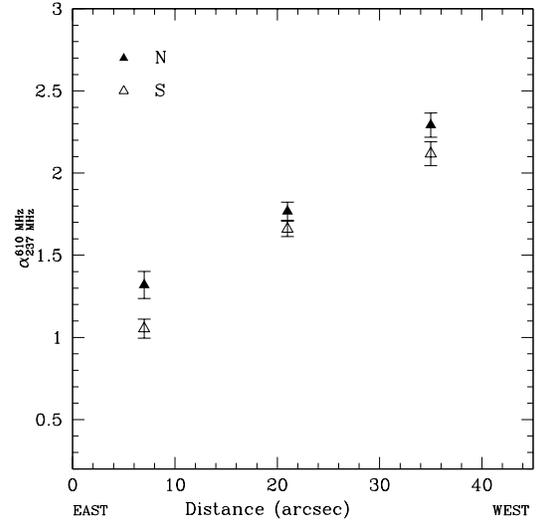}
\caption{{\it Left} -- Grids used to derive the spectral 
index trend overlaid on the 327--610 MHz spectral index image 
(grey scale; same as Fig.~\ref{fig:sp_images}). {\it Right} -- 
Trend of the spectral index $\alpha^{610}_{327}$ in the northern
(filled triangles) and southern (empty triangles) parts of the relic.}
\label{fig:a521_steep}
\end{figure*}
%
%

In order to check the significance of the spectral
steepening visible in Fig.~\ref{fig:sp_images}, we 
determined the average spectral index in 3 independent 
strips of $14^{\prime \prime}\times80^{\prime \prime}$ size 
in the northern region of the relic and 3 independent
strips of $14^{\prime \prime} \times 100^{\prime \prime}$ size 
in the southern part. These two regions are labelled respectively 
N and S in the left panel of Fig.~\ref{fig:a521_steep}. 
The central portion of the relic was excluded from the analysis 
because of the presence of the radio galaxies B and C 
(Fig.~\ref{fig:a521_steep}) which may affect the spectral 
trend. The strips were set parallel to the edge of the relic, 
i.e., at a position angle of $5^{\circ}$ and $32^{\circ}$ in 
the N and S regions, respectively. For each strip we integrated the 
flux density on the 327 MHz and 610 MHz images individually, 
and then calculated the corresponding spectral index values. 
The average 327--610 MHz spectral index in each strip is 
shown in the right panel of Fig.~\ref{fig:a521_steep}. 
The spectral index trend shows a gradual steepening going 
westwards both in the N and S regions. In particular 
$\alpha$ ranges from $\simeq 1.0\pm 0.1$ to $\simeq 
2.1\pm 0.1$ in the southern part (empty triangles), and 
from $\simeq 1.3 \pm 0.2$ to $\simeq  2.3 \pm 0.1$ across 
the northern region (filled triangles).
\\
\\
A steepening from the outer to the inner edge of the 
relic with available spectral index images has been 
observed also in few other cases, as for example 
in A\,3667 (R\"ottgering et al. 1997) and A\,2744 
(Orr\'u et al. 2007). A more detailed study of the spectral 
index gradient, similar to the analysis presented in this paper, 
has been carried out for the relic in A\,2256 by 
Clarke \& En\ss{}lin (2006), who found a significant 
steepening of $\alpha$ between 1369 MHz and 1703 MHz from 
the external edge of the relic toward the cluster core, i.e., 
the same kind of gradient found in the A\,521 relic.

\section{Integrated radio spectrum of the relic}\label{sec:spectrum}

The wealth of multifrequency radio observations available for the 
relic in A\,521 allows the determination of its integrated spectrum 
over almost two orders of magnitude in frequency. 
\\
In Tab.~\ref{tab:a521_relic} we report all the available flux 
densities of the relic, along with the resolution of the images 
used for the measurement\footnote{The table provides also
the source flux density at 235 MHz measured on a preliminary 
image from very recent GMRT observations (see Appendix).}. 
In order to obtain a consistent measurement of the total flux 
densities, we integrated over the same region in all 
the images, and subtracted the flux density of the embedded point--sources, 
as measured on the corresponding full resolution images. 
The source subtraction could not be applied for the 74 MHz value,
since the low resolution (80$^{\prime \prime} \times 80^{\prime \prime}$)
of the Very Low--frequency Sky Survey (VLSS\footnote{http:$//$lwa.nrl.navy.mil$/$VLSS$/$}) 
image does not allow to perform such operation. Furthermore the extended 
structure visible in the VLSS image is not fully consistent with the relic 
morphology at higher frequencies, probably due to the different angular resolution
and the much lower sensitivity of the VLSS (average rms noise of the order
of 1$\sigma$=0.1 Jy b$^{-1}$). For these reasons the flux density 
measurement at 74 MHz is very uncertain.

%
%
\begin{table}[h!] 
\caption[]{Flux density values of the relic.}
\begin{center}
\begin{tabular}{lccc}
\hline\noalign{\smallskip}
$\nu$ & S$_{\nu}$ & HPWB & Ref. \\
 (MHz) & (mJy) & $^{\prime \prime} \times^{\prime \prime}$ & \\
\hline\noalign{\smallskip}
74  & 660 & 80.0$\times$80.0  & VLSS \\
235 & 180$\pm$10 & 15.6$\times$12.4 & see Appendix \\
327 & 114$\pm$6  & 16.0$\times$13.0  & this work; Fig.~\ref{fig:relic327} \\
610 & 42$\pm$2    & 13.0$\times$8.1   & GVB06  \\
1410 & 14$\pm$1   & 15.0$\times$12.0  & Giacintucci (2007) $^{\star}$ \\
4890 & 2.0$\pm$0.2 & 22.0$\times$15.0 & this work; Fig.~\ref{fig:a521_relic4.9} \\
\hline\noalign{\smallskip}
\end{tabular}
\end{center}
$\star$ from VLA--CnB archival data (Obs. Id. AF\,0390).
\label{tab:a521_relic}
\end{table}

\begin{figure}[h!]
\centering
\includegraphics[angle=0,width=8cm]{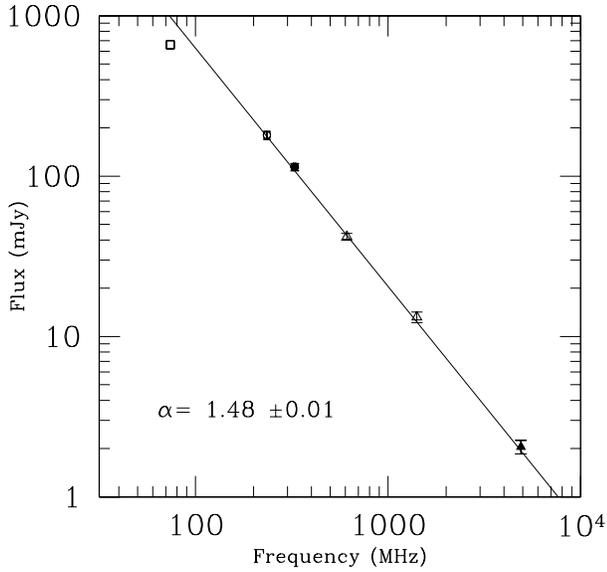}
\caption{Radio spectrum of the radio relic between 74 MHz
and 4.9 GHz. The solid circle is the relic flux density
measured on the new GMRT image at 327 MHz. The empty triangles 
are the 610 MHz data point from GVB06, and the 1.4 GHz value 
from VLA archival data. The solid triangle is the source flux
density measured on the new VLA image at 4.9 GHz. The empty
circle is the flux density from a preliminary GMRT image at 235 MHz
(see Appendix). The empty square is the 74 MHz flux density from the 
VLSS image. The solid line represents the linear fit to the data
by using the points between 327 MHz and 4.9 GHz.}
\label{fig:a521_spectrum}
\end{figure}

The integrated spectrum of the relic is shown in 
Fig.~\ref{fig:a521_spectrum}. The spectrum does not 
show any steepening up to 4.9 GHz and is well fitted by 
a single power law with slope 
$\alpha=1.48 \pm 0.01$ (solid line) between 
235 MHz and 4.9 GHz. Given its large uncertainty, the 74 MHz 
data point was not included in the fit. We note that the 
4.9 GHz flux density should be 
considered a lower limit, given the array, u--v coverage 
and resolution which led to its detection (see Sec.~4.2). 

New deep observations carried out with 
the VLA at 74 (October 2007) will allow us to better 
constrain the low frequency end of the radio spectrum.

\section{Discussion}\label{sec:disc}

A preliminary discussion on the formation of the radio 
relic in A\,521 was carried out in GVB06, where a 
number of possible theoretical frameworks were taken 
into account, all related to the assessed ongoing merging 
activity in this cluster. 
Two possible scenarios were considered, both 
invoking a tight connection with the presence of a 
merger--driven shock front at the location of the 
relic. Such a shock may accelerate electrons 
to ultra--relativistic energies (En\ss{}lin et al. 
\cite{ensslin98}; Roettiger et al. 1999; Hoeft \& Br\"uggen 2007), 
or it may {\it revive} fossil radio plasma through adiabatic 
compression of the magnetic field (En\ss{}lin \& 
Gopal--Krishna \cite{ensslin01}).

A third alternative scenario was also proposed, based on 
the hypothesis of a physical link between the diffuse 
emission of the relic and the nearby radio galaxy 
J\,0454--1016a, as suggested by the faint radio bridge 
of emission between the two sources observed at 610 MHz
(e.g., Fig.\ref{fig:a521_relic4.9}, right panel). 
The high frequency images of J\,0454--1016a 
(Fig.~\ref{fig:a521_compo}) and the spectral analysis of the 
radio relic presented in this paper (Secs.~\ref{sec:spix}
and \ref{sec:spectrum}) allow us to carry out a more 
detailed comparison between the expectations from 
the models and the observed properties of 
the relic.

%
%
%
%

\subsection{Connection of J0454--1016a with the radio relic}

GVB06 proposed that the relic might be the result of 
the ram pressure stripping of the radio lobes of 
J\,0454$-$1016a by either {\it i)} group merging activity 
in the southern cluster region, or {\it ii)} the infall 
of the radio galaxy itself into the cluster. On the basis 
of pressure arguments, and given the projected distance 
of J\,0454$-$1016a from the relic, they estimated that 
the infall velocity of the merging group, or of the 
galaxy itself, should be $\gtsim$~3000 km s$^{-1}$ 
(leading to a shock with Mach number $M ~ \gtsim ~ 2$) 
to allow the electrons in the radio lobes to still emit 
in the radio band. 

The high resolution and high frequency observations presented 
in Sec.~\ref{sec:a521_j0454} do not seem to support this scenario, 
since there is no obvious morphological link between the relic and 
J\,0454$-$1016a. The radio galaxy has a head--tail morphology, 
whose tails extend (in projection) in a direction nearly opposite  
to the relic region and to the faint brigde of emission 
connecting the galaxy and the relic (Fig.~\ref{fig:a521_compo}). 
Even though the origin of the faint bridge remains unclear, it seems 
to be unrelated to the current AGN activity of J\,0454$-$1016a.

%
%
\begin{figure*}[t]
\centering
\includegraphics[angle=0,width=8.5cm,height=8.4cm]{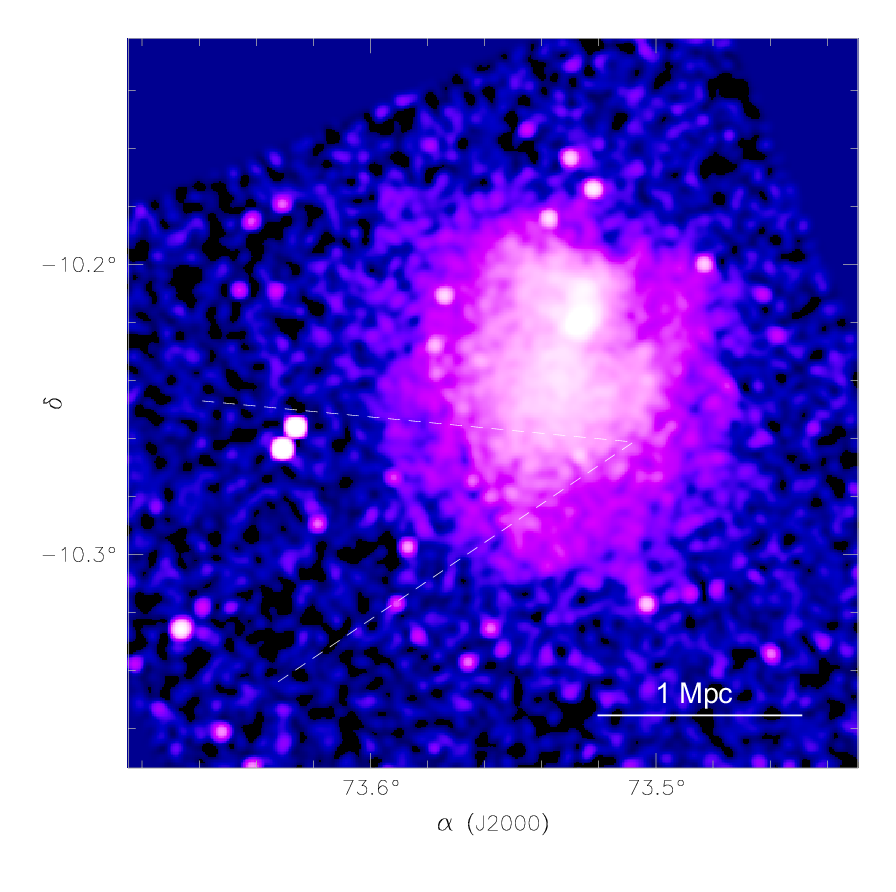}
\hspace{0.5cm}
\includegraphics[angle=0,width=8.6cm,height=8.5cm]{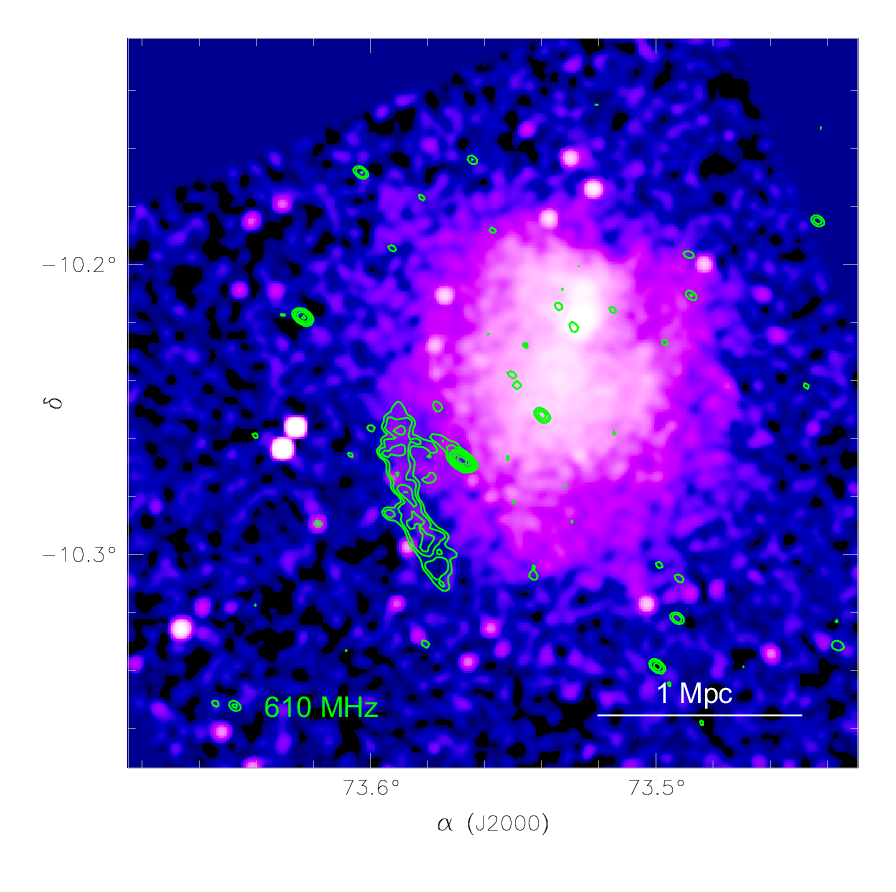}

\caption{{\it Left panel} -- Smoothed {\it Chandra} image of A\,521 
in the 0.5--4 keV energy band. {\it Right Panel} -- Same image as left 
panel with the GMRT 610 MHz contours of the A\,521 field overlaid. 
The resolution of the radio image is 13.1$^{\prime \prime} \times 
8.1^{\prime \prime}$, p.a. 56$^{\circ}$. Contour levels are spaced by a 
factor 2 starting from 0.2 mJy b$^{-1}$.} 
\label{fig:chandra}
\end{figure*}
%
%

\subsection{Basic expectations from the shock scenarios 
and comparison with the observations}

The shock scenarios for the origin of radio relics provide 
a number of expectations which can be tested by means of
the comparison with the observed properties of the source. 
\\
\\
{\it a) Shock acceleration} -- In this scenario
the relic emission is produced by electrons accelerated 
from the thermal gas to relativistic energies by the passage 
of a merger shock wave (En\ss{}lin et al. \cite{ensslin98}; 
Roettiger et al. 1999; Hoeft \& Br\"uggen 2007). 
Assuming linear shock acceleration theory, in the case of a fully 
ionised plasma 
as the ICM, the steady state energy spectrum of the electrons 
in the relic is a power law whose slope $\delta$ is related 
to the shock Mach number $M$ according to the following 
equation (e.g., Blandford \& Eichler 1987):

\begin{equation}\label{eq:mach}
\delta = 2 \frac{M^2+1}{M^2-1}+1 
\end{equation}

\noindent where the effect of particle aging $\Delta(\delta)=1$ 
(from the combined effect of inverse Compton energy losses and 
continous injection) is included (Sarazin 1999). The ensuing integrated
radio emission of the relic has a single power law spectrum 
with $\alpha=(\delta -1)/2$, which is thus related to the Mach 
number of the shock through Eq.~\ref{eq:mach}. More specifically,
the total spectrum results from the combination of
the aged spectrum of the injected electrons: once accelerated, 
the electrons are short--lived because of the inverse Compton and 
synchrotron energy losses, and their spectrum will rapidly 
steepen with the distance from the shock rim. Thus, the 
spectral index distribution of the relic synchrotron radio emission 
would exhibit a progressive steepening going from the current location 
of the shock (where the observed $\alpha$ is the injection spectral 
index) to the trailing edge.
\\
\\
\noindent {\it b) Adiabatic compression} -- In this scenario
the relic is the result of the adiabatic compression exerted 
by a merger shock on a region containing fossil radio plasma 
(En\ss{}lin \& Gopal--Krishna \cite{ensslin01}). Such 
compression might increase the magnetic field strength and 
re--energize the electron population in the fossil plasma, 
thus leading to observable radio emission. In this case the 
relativistic electrons are expected to produce diffuse radio 
emission in front of the bow shock, and then rapidly lose 
their energy while moving away from the front. Hence a 
transversal steepening in the spectral index 
distribution of the relic may be expected also in this model. 
Unlike the shock acceleration case, this scenario predicts 
a curved integrated radio spectrum showing a spectral 
steepening at high frequency, at least for relatively weak
shocks ($M$~$\ltsim$~3).
\\
\\
\noindent {\it c) Shock re--acceleration} -- This
scenario was discussed by Markevitch et al. (2005) 
for the origin of the radio edge coincident with the 
shock front in A\,520. In this case the relic is produced by 
re--acceleration of fossil electrons in the ICM by the shock. 
The basic expectations for the spectral properties of
the relic are similar to the shock acceleration scenario
({\it a}). 
\\
\\
The results of the spectral analysis presented in this 
paper provide insightful information to discriminate 
between the models described above. In the case 
of the relic in A\,521, a significant steepening of
the spectral index was found going from the eastern 
edge of the source toward the western border (Fig.~7). 
This behaviour is in line with the expectations of all
the scenarios. It implies that the shock is expected
to be moving outwards with respect to the cluster 
centre, and its current location should be approximately 
coincident with the eastern edge of the relic. 
In Sec.~\ref{sec:spectrum} we showed that 
the integrated spectrum of the relic is well reproduced 
by a single power law with steep spectral index 
($\alpha \sim 1.5$), and with no evidence of high frequency 
steepening up to 4.9 GHz (Fig.~\ref{fig:a521_spectrum}). 
Such a spectral shape is expected only in 
the framework of the shock acceleration and re--acceleration 
scenario. 
\\
The adiabatic compression scenario should
produce curved spectra with a high frequency cut--off in 
the case of moderate or weak shocks, such as those expected 
to be developed during cluster mergers (e.g., 
Gabici \& Blasi 2003; Ryu et al. 2003; Pfrommer et al. 2006, 
and references therein). In principle a cut--off just above
5 GHz cannot be ruled out. The compression enhances the magnetic 
field and the particle energy density, moving the break 
frequency in the synchrotron spectrum at higher 
frequencies. Assuming thermal and relativistic particles
mixed into the fossil plasma, the ratio of the post--compression 
and initial break frequencies ($\nu_b^{post}$ and $\nu_b^{0}$, 
respectively) is (e.g., Markevitch et al. 2005): 

\begin{equation}\label{eq:breakfreq}
\frac{\nu_b^{post}}{\nu_b^{0}} = \left( \frac{4M^2}{M^2 +3} \right)^{4/3} < 6
\end{equation}

\noindent This value is not high enough to explain the lack of 
a cut--off in the observed spectrum of the relic\footnote{Adopting a viable
value of $\nu_b^{0}$=50--100 MHz in a fossil radio plasma.}. 
\\
This factor can be strongly increased if the shock 
compression acts on a ghost of purely relativistic plasma (i.e., 
not mixed with the thermal ICM). In this case (En\ss{}lin \& Gopal--Krishna 2001): 

\begin{equation}\label{eq:breakfreq2}
\frac{\nu_b^{post}}{\nu_b^{0}} \approx \frac{\rm P_2}{\rm P_1} \approx \frac{5M^2-1}{4} 
\end{equation}

\noindent where $\rm P_1$ and $\rm P_2$ are the pre-- and 
post--shock thermal pressures. To produce a spectral break 
at $\nu \ge 5$ GHz, a shock with $M \ge 7$ is required, which however 
would imply a very unlikely pre--shock temperature of T$<1$ keV 
for A\,521. 
\\
An additional point is that the injection spectrum of the fossil
radio plasma in this scenario should be roughly equal to the
observed one ($\alpha \sim 1.5$), which however is much steeper
than typical spectra of radio galaxies.
\\
\\
To summarise, our spectral analysis suggests that the
origin of the relic in A\,521 is consistent with the
shock acceleration scenario. According to Eq.~\ref{eq:mach} 
we can estimate the Mach number of the shock responsible 
for the electron acceleration. The spectral fit provided a 
total spectral index $\alpha=1.48 \pm 0.01$ (Sec.~\ref{sec:spectrum}), 
which corresponds to $\delta=3.96 \pm 0.02$, and thus
$M(\delta)= 2.27 \mp 0.02$. Such Mach number is 
in reasonable agreement with the values expected 
for the cluster merger shocks, and indeed 
observed in merging clusters (e.g., Markevitch \& Vikhlinin 
2007). As argued in GVB06, the relic is located in a 
pheripheral region of A\,521 which is expected 
to be dynamically active (Maurogordato et al. 
2000; Ferrari et al. 2003 and 2006; see also 
Fig.~2 in GVB06). Thus the presence of a shock 
front at the relic location is likely.

%
%
\begin{figure*}[t]
\centering
\includegraphics[angle=0,width=8cm]{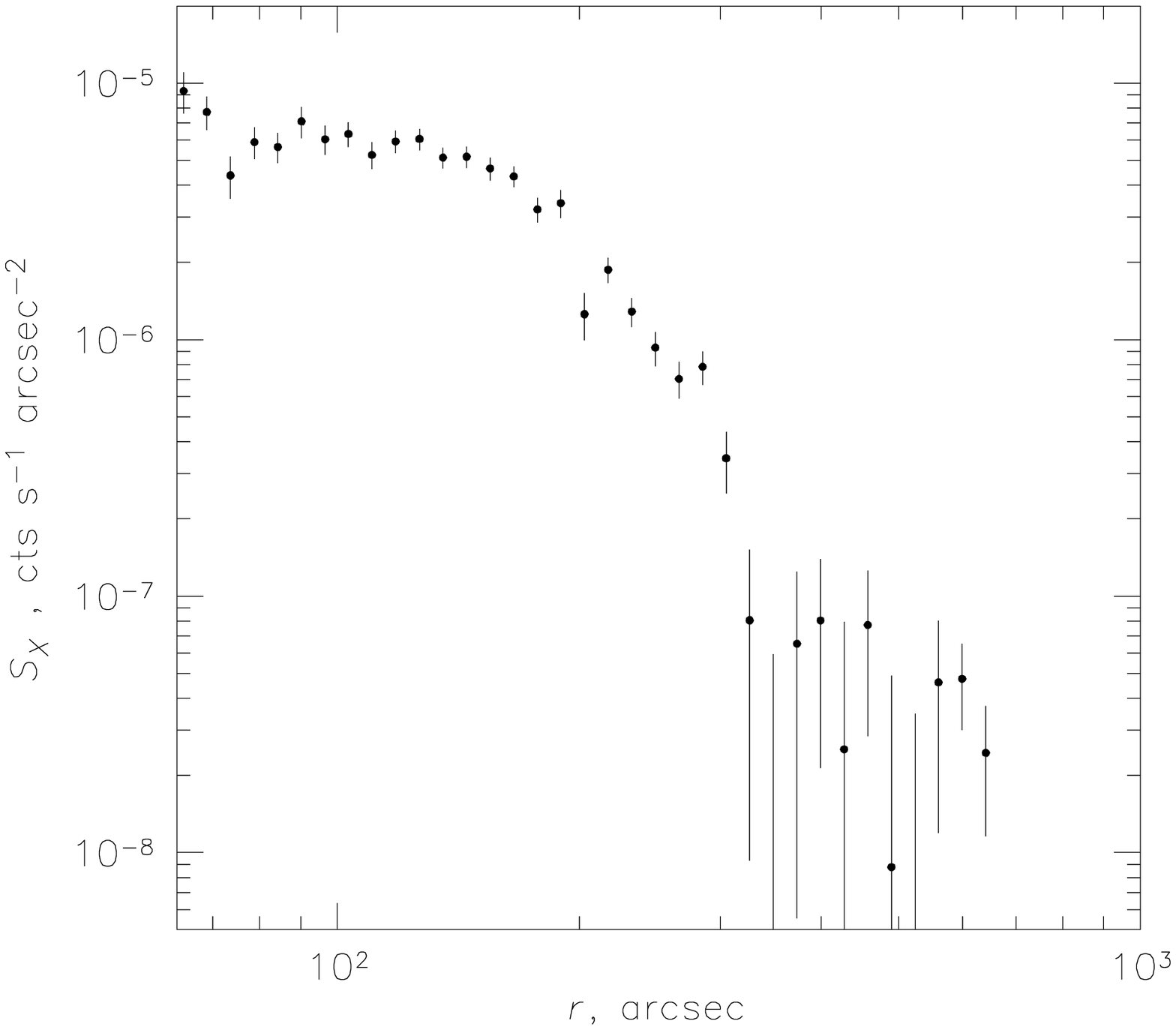}
\hspace{0.5cm}
\includegraphics[angle=0,width=8.7cm]{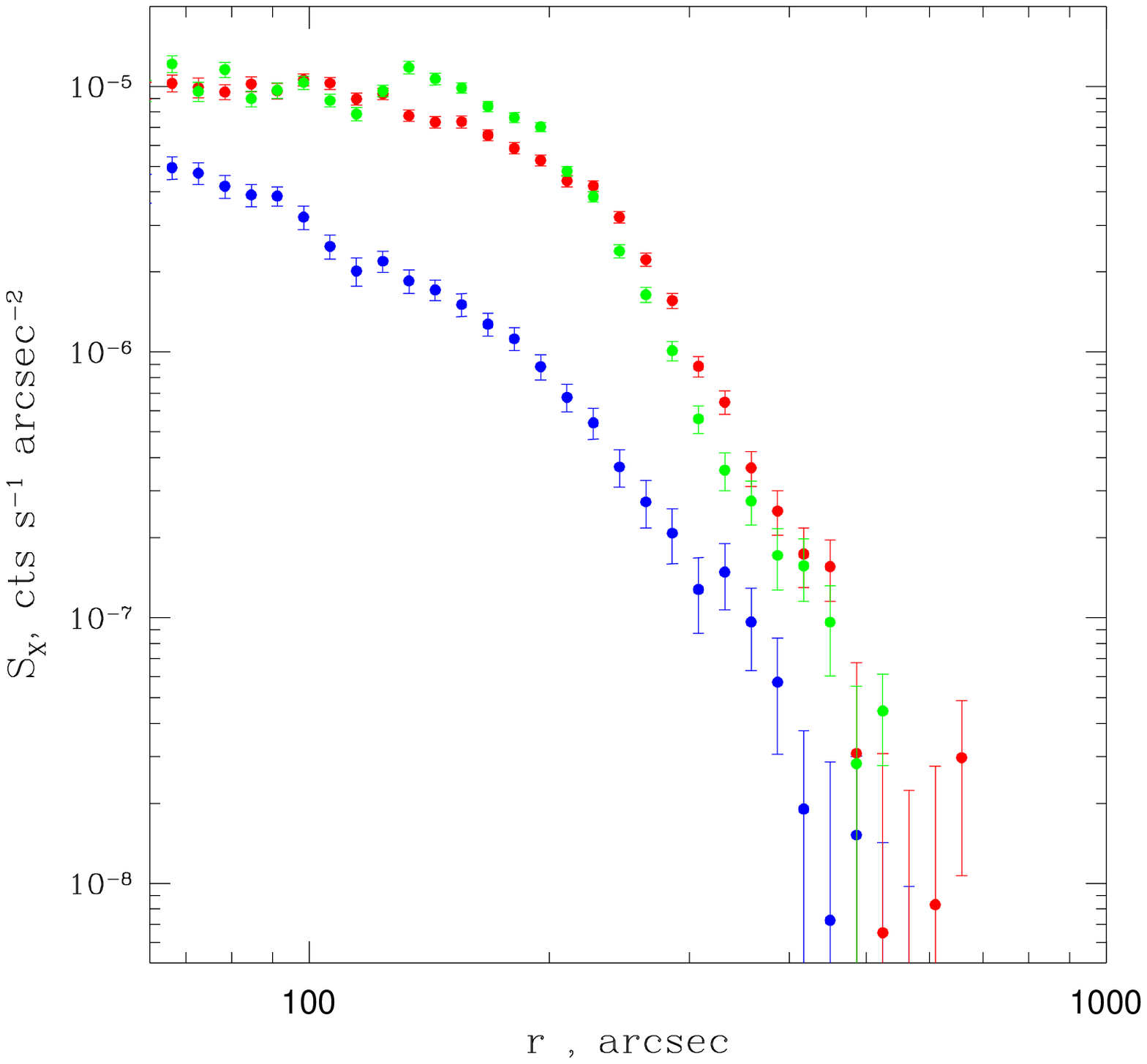}
\caption{{\it Left panel:} X--ray 0.5--4.0 keV brighntess profile across the edge in
the relic region, extracted in the sector indicated in the left panel of 
Fig.~\ref{fig:chandra}. Error bars are $1\sigma$. {\it Right panel:} X--ray 
0.5--4.0 keV brightness profiles extracted in sectors covering the South--West 
(blue), North--East (red) and North--West (green) portions of the cluster.
Error bars are $1\sigma$.}
\label{fig:xprofile}
\end{figure*}
%
%

%
%
\begin{figure}[h!]
\centering
\includegraphics[angle=0,width=8cm]{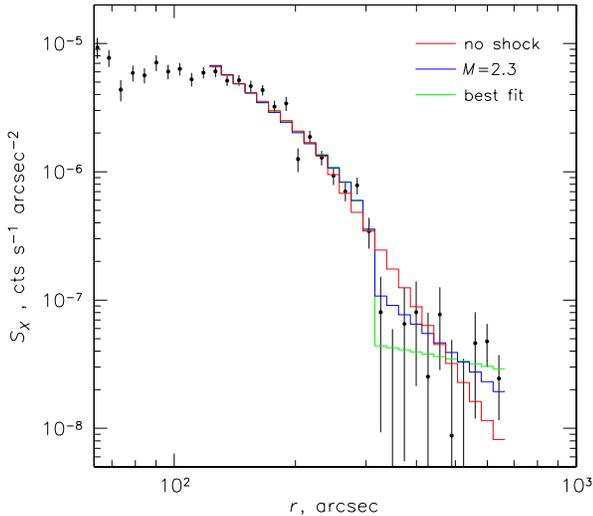}
\caption{X--ray brighntess profile (same as left panel of 
Fig.~10) compared to different models. The green and red lines 
are the fit with and without a shock discontinuity, respectively. 
The blue line is the profile expected for a shock with $M=2.3$.}
\label{fig:xprofile2}
\end{figure}
%
%

\subsection{A shock front coincident with the radio relic?}

Using the available {\it Chandra} X--ray archivale data of A\,521, 
we checked if there is indeed a shock front in the cluster gas 
at the position of the relic. Shock fronts in clusters are 
observed very rarely. Only two unambigous examples, exhibiting 
both a sharp gas density edge and a clear gas temperature jump, 
have been discovered by {\it Chandra} so far, those in 
the merging clusters 1E\,0657--56 (Markevitch et al. 2002) 
and A\,520 (Markevitch et al. 2005). Their rarity is due to the 
fact that the viewing geometry and the moment of our observation 
must be very favourable: a merger shock quickly moves to the cluster 
outskirts where it cannot be detected. 

For A\,521 two observations are available in the {\it Chandra} 
public archive, one performed with ACIS--I and another with 
ACIS--S, with $\sim$40 ks exposure each (Ferrari et al. 2006). 
Unfortunately, the relic lies right at the S3 chip boundary in the 
ACIS--S observation, so we could use only the ACIS--I observation 
(OBSID 901) for our purpose. We cleaned the data and modeled the 
detector background and instrumental spatial response as described 
most recently in Vikhlinin et al. (2005). 

In Fig.~\ref{fig:chandra} we show the resulting {\it Chandra} image in 
the 0.5--4.0 keV energy range. The background was subtracted and the
image was divided by the exposure map, and then smoothed with a 
$\sigma=6^{\prime \prime}$ Gaussian. In the right panel the 
GMRT 610 MHz contours are overlaid on the same X--ray image. The image
reveals a clear brightness edge coincident with the outer 
edge of the radio relic. 
\\
The radial brightness profile in Fig.~\ref{fig:xprofile} (left panel) 
shows this edge more clearly. The profile was extracted from the unsmoothed 
image in the sector shown in the left panel of Fig.~\ref{fig:chandra}, which 
is centred on the centre of curvature of the relic and spans the angle covered 
by the source. Discrete X--ray sources were excluded for the profile
derivation. Fig.~\ref{fig:xprofile} (left panel) shows a  
brightness edge at r=300$^{\prime \prime}$, which coincides
with the outer edge of the radio relic (the drop in the profile 
and increased error bar at r$\approx 200^{\prime \prime}$ 
corresponds to an interchip gap). For comparison, in the right 
panel we plot the X--ray brightness profiles extracted from 
three other sectors of the cluster. Neither of them shows such 
an edge, consistently with the visual inspection of the image
shown in Fig.\ref{fig:chandra}.

The X--ray edge in Fig.~\ref{fig:xprofile} (left panel) has
the characteristic shape that corresponds to a projection of
a spherical density discontinuity. To quantify this
discontinuity, we tried to fit this brightness profile in
the immediate vicinity of the edge with a gas density model
consisting of two power laws and an abrupt jump, with power
laws and the position and amplitude of the jump being free
parameters (as in Markevitch et al.\ 2000). A continuous
density profile (i.e., no jump, but a possible break), shown
by the red line in Fig.~\ref{fig:xprofile2}, is inconsistent
with the data at about $4.5\sigma$.  If one assumes a 6 keV
gas temperature just inside the edge (Ferrari et al.\ 2006,
determined from this and the ACIS-S {\it Chandra}\/
observations) and converts brightness in the {\it Chandra} 0.5--4
keV band into plasma emission measure self-consistently, the
best-fit density jump (the model shown by green line)
corresponds to $M\approx 7$. However, as mentioned above,
such a shock would imply a pre-shock temperature of $\sim
0.4$ keV, which is very unlikely close to the center of such a
hot cluster. A shock with $M=2.3$, predicted from the radio
spectrum (Sec.~7.2), corresponds to an edge shown by the
blue line in Fig.~\ref{fig:xprofile2}; this fit is only
$1.8\sigma$ away from the best fit, which we consider a good
agreement.

Unfortunately, the accuracy of the existing {\it Chandra}\/
observation is not sufficient to measure the gas temperature
on the faint side of the edge. Thus, we cannot rule out
other interpretations for this feature, e.g., a cold front
(Markevitch \& Vikhlinin 2007). However, a cold front with
such a density contrast would correspond to an outer gas
temperature in excess of 15 keV --- quite unlikely in a
cluster of this X-ray luminosity. We conclude that the X-ray
edge is most likely a shock front --- found at a location
and with an amplitude exactly as needed to produce the radio
relic via the shock acceleration mechanism.

\section{Summary and Conclusions}\label{sec:summary}

We presented new deep radio images of the radio relic 
in A\,521 obtained using the GMRT at 327 MHz and the 
VLA at 4.9 and 8.5 GHz. We performed a study of the 
spectral properties of the source by combining the new 
radio information with our previous GMRT data at 610 
MHz (GVB06) and 1.4 GHz observations available in the 
VLA public archive and re--analysid in Giacintucci (2007).
\\
In the following we summarise the main results of our 
analysis:

\begin{itemize}

\item[1.] The relic morphology at 327 MHz is in overall
agreement with the structure observed at 610 MHz 
and 1.4 GHz. 

\item[2.] The relic was detected at 4.9 GHz at 
a high level of significance (from $3\sigma$ in the
fainter regions to $12 \sigma$ at the peaks). 
This is the second detection of a radio relic
at a frequency as high as 4.9 GHz after the
relic source 1253+275 in the Coma cluster
(Thierbach, Klein \& Wielebinski \cite{thierbac03}).

\item[3.] A faint bridge of emission is clearly
detected between the northern part of the relic
and the cluster radio galaxy J0454--1016a. This
emission was also observed at 610 MHz, and 
suggested the existence of a physical connection 
between the two sources. The high resolution 
images at 4.9 and 8.5 GHz seem to rule out such
hypothesis: J0454--1016a has a head--tail structure 
extending in a direction opposite with respect 
to the bridge, whose origin remains thus unclear. 

\item[4.] The spectral index image 
of the relic between 327 and 610 MHz looks rather 
patchy and irregular. 
However our analysis revealed a significant overall
gradual steepening of the spectral index along
the source minor axis, from the outer edge 
of the relic toward the cluster centre. A 
similar trend has been observed in few other 
relics with available spectral index images, 
as for example the case of A\,2256 where a 
spectral study similar to the analysis 
presented in this paper was performed 
by Clarke \& En\ss{}lin (2006).

\item[5.] We derived the integrated 
spectrum of the relic over almost two orders 
of magnitude in frequency. Such a frequency 
coverage is available for very few relics 
only (e.g., the Coma cluster relic; Thierbach, 
Klein \& Wielebinski \cite{thierbac03}). 
The spectrum of the relic 
in A\,521 is well fitted by a power law with a 
steep spectral index ($\alpha \sim 1.5$),
with no evidence of a steepening in the high 
frequency regime (up to 4.9 GHz). 
This situation is similar to the Coma relic, 
which shows a single power law spectrum with 
$\alpha \sim 1.2$.

\item[6.] We analysed the archival {\it Chandra} 
observation of A\,521 and discovered a clear X--ray 
brightness edge at the position of the outer border 
of the radio relic. This edge can be the shock 
front responsible for the electron acceleration. 
The present X--ray data accuracy is insufficient to 
confirm that it is a shock front, but the 
coincidence is tantalizing.

\end{itemize}

The results of our spectral analysis were discussed 
in the framework of the possible models for the relic origin 
considered in GVB06. We concluded that the more plausible 
scenario is the shock acceleration, while adiabatic compression
seems to be ruled out by the present data. In the shock 
acceleration case, the flattest spectrum emission of the relic 
(i.e., the outer region) 
would mark the position of the shock front, where the electron 
acceleration is expected to be currently ongoing. 
This suggests that the shock wave is propagating from the 
north--western direction toward South--East (projected 
on the plane of the sky). The possible merging scenario 
for A\,521 (Ferrari et al. 2003 and 2006) might be 
consistent with such hypothesis. The presence of an 
edge in the X--ray surface brightness in the relic region
is suggestive of the existence of a shock front which 
may be responsible for the electron acceleration in the 
relic. Deeper X--ray observations are needed to confirm 
the presence of such a shock.
\\
\\
{\it Acknowledgements.}
We thank the staff of the GMRT for their help during the observations.
GMRT is run by the National Centre for Radio Astrophysics of the Tata 
Institute of Fundamental Research. Thanks are due to E. Fomalont for his
help with the use of the NRAO pipeline. We thank the anonymous referee
for useful comments. S.G. and T.V. acknowledge partial support from the 
Italian Ministry of Foreing Affairs. This work has been partially 
supported by contracts ASI--INAF I/088/06/0, PRIN--MUR 2005, and 
PRIN--INAF 2005.

\appendix
\section{Preliminary results of the GMRT follow up at 235 MHz}
%
%
\begin{figure}[h]
\centering
\includegraphics[angle=0,width=\hsize]{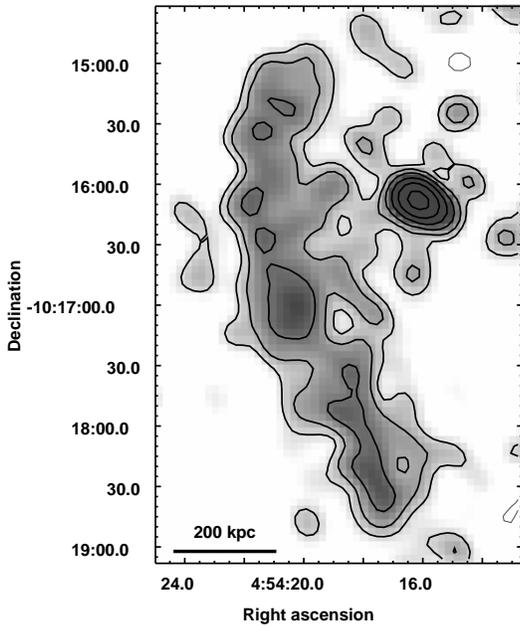}
\caption{Preliminary GMRT image (contours and grey scale) of radio 
relic at 235 MHz. The resolution is 15.6$^{\prime \prime}$ $\times$ 
12.4$^{\prime \prime}$, in p.a. 56. The noise level is 
1$\sigma$=270 $\mu$J b$^{-1}$. Contours are spaced by a factor 2
starting from $\pm 3 \sigma$. } 
\label{fig:relic235}
\end{figure}
%
%
In this Appendix we present the preliminary 235 MHz 
image of the relic obtained from very recent GMRT data. The
source was observed on 2007, November 11th and 13rd, for a
total integration time of $\sim$18 hours. The observations
were carried out in spectral line mode using only the USB 
for a total bandwidth of 8 MHz (128 channels, spectral
resolution 62.5 kHz/channel). The data reduction was 
performed as described in Sec.~2.1.  
\\
The full 
resolution image shown in Fig.~\ref{fig:relic235} was 
produced using only the data from the first day 
(i.e., $\sim$ 9 hours). The image has a very high 
sensitivity level (1$\sigma$=270 $\mu$Jy b$^{-1}$), and 
the relic is clearly detected and imaged in its whole 
extent. Given the high quality of this image we are 
confident that the flux density given in Tab.~4 is 
definitely reliable, with an error of the order of 5\%.
The analysis of the entire dataset (i.e., from the 
combination of the two observing days) will be presented 
in a forthcoming paper.

\end{document}